\documentclass[12pt]{article}

\usepackage[numbers,sort&compress]{natbib}
\usepackage{graphicx}
\usepackage{amsmath}
\usepackage{amssymb}
\usepackage{hyperref}
\usepackage{feynmp}

\newcommand{\beq}{\begin{equation}}
\newcommand{\eeq}{\end{equation}}
\newcommand{\beqs}{\begin{eqnarray}}
\newcommand{\eeqs}{\end{eqnarray}}
\newcommand{\Tr}{\ensuremath{\mathop{\mathrm{Tr}}}}

\def\ni{\noindent}
\def\be{\begin{equation}}
\def\ee{\end{equation}}
\def\bea{\begin{eqnarray}}
\def\eea{\end{eqnarray}}
\def\bsp{\be\begin{split}}
\def\la{\langle}
\def\ra{\rangle}
\def\dag{\dagger}

\def\G{\Gamma}

\def\a{\alpha}
\def\b{\beta}
\def\g{\gamma}

\def\d{\delta}
\def\e{\epsilon}
\def\m{\mu}

\def\n{\nu}

\def\l{\lambda}
\def\t{\tau}

\def\p{\partial}

\def\bR {\mathbb{R}}

\makeatletter

\newcommand{\Rmnum}[1]{\expandafter\@slowromancap\romannumeral #1@}
\makeatother

\renewcommand{\title}[1]{\vbox{\center\LARGE{#1}}\vspace{5mm}}
\renewcommand{\author}[1]{\vbox{\center\large{#1}}\vspace{5mm}}
\newcommand{\address}[1]{\vbox{\center\em#1}}
\newcommand{\email}[1]{\vbox{\center\tt#1}\vspace{5mm}}

\begin{document}
\bibliographystyle{utphys}

\begin{fmffile}{graph1}
\fmfset{zigzag_len}{12}
\fmfset{zigzag_width}{2}
\fmfset{arrow_len}{8}
\fmfset{arrow_ang}{20}

\begin{titlepage}
\hfill {\tt HU-EP-10/41}\\
\title{\vspace{1.0in} {\bf Wilson Loops in ${\cal N}=2$ Superconformal
    Yang-Mills Theory}}

\author{Roman Andree$^1$ and Donovan Young$^2$}

\address{Humboldt-Universit\"at zu Berlin, Institut f\"ur Physik,\\
  Newtonstra\ss e 15, D-12489 Berlin, Germany }

\email{$^1$randree, $^2$dyoung@physik.hu-berlin.de}

\abstract{\ni We present a three-loop (${\cal O}(g^6)$) calculation of
  the difference between the expectation values of Wilson loops
  evaluated in ${\cal N}=4$ and superconformal ${\cal N}=2$
  supersymmetric Yang-Mills theory with gauge group $SU(N)$ using
  dimensional reduction. We find a massive reduction of required
  Feynman diagrams, leaving only certain two-matter-loop corrections
  to the gauge field and associated scalar propagator. This
  ``diagrammatic difference'' leaves a finite result proportional to
  the bare propagators and allows the recovery of the $\zeta(3)$ term
  coming from the matrix model for the 1/2 BPS circular Wilson loop in
  the ${\cal N}=2$ theory. The result is valid also for closed Wilson
  loops of general shape. Comments are made concerning light-like
  polygons and supersymmetric loops in the plane and on $S^2$.}

\end{titlepage}

\tableofcontents
\section{Introduction and results}

The study of supersymmetric Wilson loops has enjoyed exciting
development since the very early days of AdS/CFT, when the basic
object and string dual were identified \cite{Maldacena:1998im,
  Rey:1998bq}. Standing at the forefront of these investigations has
been the 1/2 BPS circle of ${\cal N}=4$ supersymmetric Yang-Mills
theory (SYM). This object is intimately related to the trivial 1/2 BPS
infinite line, through the singular conformal inversion $x^\mu \to
x^\mu/x^2$, which leaves only the point at infinity for the site of
non-trivial dynamics, which are therefore captured by a 0-dimensional
quantum field theory - the celebrated matrix model of
\cite{Erickson:2000af,Drukker:2000rr}. Perhaps the most potent feature
of this matrix model is that it captures three very different regimes
in the dual string theory, corresponding to the scaling of the rank
$R$ of the representation which the trace is taken in, with respect to
$N$, the rank of the gauge group. For $R \sim N^0$, one has a
semi-classical fundamental string describing a minimal surface in
$AdS_5\times S^5$, for $R\sim N^1$ the string becomes a D-brane (or
collection thereof) again in $AdS_5\times S^5$, while for $R\sim N^2$
the back-reaction of the branes deform the background geometry and
$AdS_5\times S^5$ is replaced by a new space. These are a very rich
set of phenomena, and the fact that they can be reduced to a
relatively simple 0-dimensional theory is astounding. Perhaps more
astounding is that the matrix model should also describe the full
quantum, string-loop-corrected versions of these objects.

Recently the precise way in which the 1/2 BPS Wilson loop comes to be
described by the matrix model has been understood through the
techniques of localization \cite{Pestun:2007rz}. Beyond providing a
previously lacking proof of the equivalence between the matrix model
and the Wilson loop, this work has opened up the study of Wilson loops
into exciting new avenues
\cite{Passerini:2010pr,Drukker:2009tz,Drukker:2009id,Drukker:2010jp,Alday:2009fs}. One
of the basic extensions provided by \cite{Pestun:2007rz} is to the
description of the 1/2 BPS circular loop in ${\cal N}=2$ SYM. In the
superconformal case, when $N_f=2N$ fundamental hypermultiplets are
coupled to the theory, the matrix model is modified with respect to
the ${\cal N}=4$ case by the insertion of a determinant
factor\footnote{And also instanton contributions, which will not
  concern us here.}. This contribution was worked out in detail in
\cite{Pestun:2007rz}, for the specific case of $SU(2)$, where it was
shown that the effect of the determinant factor in a perturbative
expansion was the addition of a term at ${\cal O}(g^6)$, proportional
to $\zeta(3)$. Using the explicit expression for the determinant
factor provided in \cite{Pestun:2007rz}, it is a trivial matter to
generalize the calculation for $SU(N)$, as we do in section
\ref{sec:mm}, and the additional $\zeta(3)$ term remains at ${\cal
  O}(g^6)$, albeit with a generalized coefficient.

The purpose of this paper is to recover this $\zeta(3)$ term from
perturbation theory. The technique we use is dimensional
reduction. Our strategy is to take the ``diagrammatic difference'' of
the ${\cal N}=4$ and ${\cal N}=2$ results. In so doing we can prove
that the calculations cancel up to ${\cal O}(g^4)$, in agreement with
the matrix model result. Further, a massively reduced set of Feynman
diagrams remains at ${\cal O}(g^6)$, all of which are two-loop
matter-corrections to the ${\cal N}=2$ adjoint gauge and scalar field
propagators. Of these, only two give $\zeta(3)$ contributions, and are
responsible for the exact match with the matrix model. We find a
complete cancellation of divergences, which are generically ${\cal
  O}(1/\e^2)$ where the dimension is taken as $4-2\e$. The result is
proportional to the bare gauge field and real scalar propagator, and
therefore is directly applicable to {\it any} closed\footnote{That the
  Wilson loop be closed is important for the results of section
  \ref{sec:triv}.} Wilson loop in the ${\cal N}=2$ theory of the form
\be\label{wl}
W = \frac{1}{N} \Tr P \exp \oint d\t \bigl( i \dot x^\m(\t) A_\m + |\dot
x(\t)| \Theta^I(\t) \Phi_I \bigr),\qquad I = 1,2,
\ee 
where the $\Phi_I$ are the two real adjoint scalars in the gauge
multiplet. The result may be compactly expressed in the following way

\bsp\label{mainres}
&\la W \ra_{{\cal N}=4} - \la W \ra_{{\cal N}=2} =\\ 
&~ g^6\left[\frac{12\zeta(3)}{(4\pi)^4} (N^2+1)\right]
 \frac{N^2-1}{2N} \frac{1}{2!} \oint
d\t_1 \oint d\t_2 \, 
\frac{|\dot x_1||\dot x_2|\Theta_1\cdot \Theta_2  -  \dot x_1 \cdot\dot x_2}{4\pi^2
  (x_1-x_2)^2}+{\cal O}(g^8),
\end{split}
\ee 

\ni where the bracketed expression is the dressing the propagators
receive, while the remainder of the expression is the standard
expansion of the Wilson loop to second order. 

The outline of this paper is as follows. We present the result
stemming from the matrix model for general $SU(N)$ in section
\ref{sec:mm}. In section \ref{sec:pt} we describe the structure
of the perturbation theory calculation, giving details in appendix
\ref{app:main}. Finally in section \ref{sec:scamp} we discuss the
implications of our result for other well-known Wilson loops,
including the Zarembo loops \cite{Zarembo:2002an}, the longitudes of
\cite{Drukker:2007qr}, and the light-like polygonal Wilson loop.

\section{Results from localization}
\label{sec:mm}

In this section we derive the result for the circular Wilson loop
expectation value in superconformal $SU(N)$ ${\cal N}=2$ SYM, coming
from the matrix model of Pestun \cite{Pestun:2007rz}. In particular we
are interested in the $\zeta(3)$ term occurring at ${\cal O}(g^6)$. We
take coordinates on the Cartan sub-algebra of $SU(N)$, $\vec a$, which
is an $(N-1)$-component vector, and the weights of the fundamental
representation ${\vec w}_i$, $i=1,\ldots,N$. The roots are given by
${\vec w}_{ij} \equiv {\vec w}_i - {\vec w}_j$. The Wilson loop
expectation value, excluding instanton contributions, is then given by
\bsp
\la W \ra = \frac{1}{Z}
\int_{-\infty}^{\infty} d^{N-1} \vec a\, 
\prod_{i\neq j}\left( {\vec w}_{ij}\cdot \vec a \right)\,
{\cal Z}\,
e^{-4\pi^2 \vec a^2 /g^2  }\,
\frac{1}{N} \sum_i e^{2\pi {\vec w}_i \cdot \vec a}, 
\end{split}
\ee
where $Z$ is the integral without the Wilson loop insertion
$\frac{1}{N} \sum_i e^{2\pi {\vec w}_i \cdot \vec a}$ included. The
determinant factor ${\cal Z}$ is absent in the ${\cal N}=4$ case, and
is given by
\be
{\cal Z} = \prod_{i\neq j} H(i {\vec w}_{ij} \cdot \vec a)\,
\Bigl(\prod_i H(i {\vec w}_i \cdot \vec a)\Bigr)^{-2N},
\ee
where $H(x)\equiv G(1+x)G(1-x)$, where $G(x)$ is the Barnes
G-function. We will require the perturbative expansion of ${\cal Z}$,
and it is simplest to expand its logarithm using
\be
\log H(x) = -(1+\gamma) x^2 - \sum_{n=2}^\infty \zeta(2n-1) \frac{x^{2n}}{n}.
\ee
Using the property $\sum_i {\vec w}_i = 0$, stemming from the
tracelessness of the group generators, one finds that the first
correction is quartic in $\vec a$
\be\label{logZ}
\log {\cal Z} = \zeta(3) \left[ N \sum_i ( {\vec w}_i \cdot \vec a )^4
  - \sum_{i<j}  ( {\vec w}_{ij} \cdot \vec a )^4\right] + {\cal
  O}({\vec a}^6).
\ee
Using the explicit construction of $SU(N)$ weights 
\bsp &{\vec w}_1 =
(\frac{1}{2},\frac{1}{\sqrt{12}},\ldots,\frac{1}{\sqrt{2N(N-1)}}),\\ &{\vec
  w}_k =
(0,\ldots,0,\frac{-(k-1)}{\sqrt{2k(k-1)}},\frac{1}{\sqrt{2k(k+1)}},
\ldots,\frac{1}{\sqrt{2N(N-1)}}),\\
\end{split}
\ee
(\ref{logZ}) may be further simplified to
\be
\log {\cal Z} = -\frac{3}{4} \zeta(3) \left( \vec a^2\right)^2+ {\cal
  O}({\vec a}^6).
\ee
We can then express this factor as a derivative by the coupling acting
on the quadratic action in the matrix model
\be
 \left( \vec a^2\right)^2 = \left(\frac{g^2}{4\pi^2}\right)^2 
\left[ \frac{d^2}{dq^2} e^{4 \pi^2 q\vec a^2 /g^2}\right]_{q=1}.
\ee
The first contribution of ${\cal Z}$ to the Wilson loop's expectation
value may then be expressed as
\be
\la W \ra =\left. \frac{\left(1+\a\, \p_q^2\right) 
(g^2/q)^{\frac{N^2-1}{2}}\left[ 1+ \frac{N^2-1}{8N}g^2/q + \ldots\right]}
{\left(1+\a\, \p_q^2\right) (g^2/q)^{\frac{N^2-1}{2}}}\right|_{q=1}
\ee
where $\a \equiv-(3/4) \zeta(3) (g^2/(4\pi^2))^2$, and the series in
square brackets is the expectation value of the circular Wilson loop in
${\cal N}=4$ SYM, with coupling $g^2/q$. The result is
\be\label{mmres}
\la W \ra_{{\cal N}=4} - \la W \ra_{{\cal N}=2} =
\frac{3\, \zeta(3)}{512\pi^4} \frac{(N^2-1)(N^2+1)}{N}\, g^6 + {\cal O}(g^8).
\ee
In the next section we will recover this result from perturbation theory.

\section{Perturbation theory}
\label{sec:pt}

\begin{figure}[t]
\begin{center}
\includegraphics*[bb=25 45 400 200, height=1.0in]{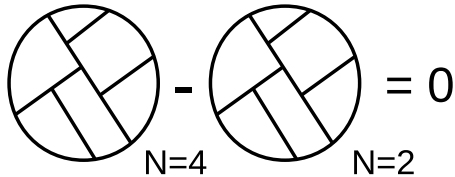}
\end{center}
\caption{``Tree'' type diagrams are identical in the two theories, and
so their difference vanishes.}
\label{fig:noloop}
\end{figure}

\begin{figure}
\begin{center}
\includegraphics*[bb=25 45 400 200, height=1.0in]{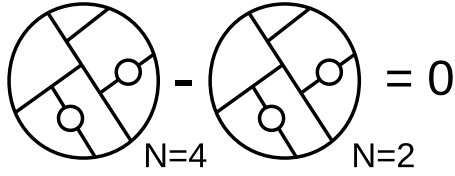}
\end{center}
\caption{One-loop corrected tree-type diagrams are also identical in
  the two theories, and so their difference also vanishes.}
\label{fig:1loop}
\end{figure}

We write the action of Euclidean ${\cal N}=2$ superconformal
Yang-Mills theory following \cite{Rey:2010ry}, as the sum of ${\cal
  N}=1$ SYM in 6-d dimensionally reduced to $4-2\e$ dimensions, and $2N$
hypermultiplets in the fundamental. In this way, one obtains the
action of ${\cal N}=4$ SYM by restricting to one adjoint
hypermultiplet as opposed to $2N$ fundamental ones. Let us write the
actions for these two theories schematically as follows (see appendix
\ref{app:main} for details)
\bsp
&S_{{\cal N}=4} = S_{{\cal N}=1}^{6\to 4-2\e} + S_{HM}^{1,\text{adj.}},\\
&S_{{\cal N}=2} = S_{{\cal N}=1}^{6\to 4-2\e} + S_{HM}^{2N,\text{fund.}}.
\end{split}
\ee
The Wilson loop under consideration does not contain couplings to the
hypermultiplet fields, it is given by (\ref{wl}) where $A_\m$ is the
gauge field, and $\Phi_I$ are the $2+2\e$ real scalar fields sitting
in $S_{{\cal N}=1}^{6\to 4-2\e}$. We now consider the difference
\be
\la W \ra_{{\cal N}=4} - \la W \ra_{{\cal N}=2}.
\ee
Let us begin at ${\cal O}(g^2)$. The only diagram is a single
gauge-field or scalar exchange. It is clear that the hypermultiplets
play no r\^ole. Therefore the difference at this order in perturbation
theory is identically zero. We can generalize this logic in the
following way. Since the ``source'' fields, i.e. those coupled in the
Wilson loop are common between the two theories, all diagrams which do
not contain loops vanish identically in the difference\footnote{This
  is because the couplings are at least quadratic in the
  hypermultiplet fields, see (\ref{n4act}), (\ref{n2act}).}, see figure
\ref{fig:noloop}.
Now let us consider the diagrams at ${\cal O}(g^4)$. There are three:
the two-rung diagram, the trivalent graph consisting of a single cubic
vertex with all three fields attached to the Wilson loop, and the
one-loop-corrected one-rung diagram. By the logic just expounded upon,
only the last diagram has a chance of surviving the difference. As we
will now show, it too cancels-out. The colour factor in the one-loop
correction to the gauge field $A_\m$ (or real scalar $\Phi_I$) propagator
stemming from a loop of one adjoint field, or $2N$ fundamental fields
is the same
\bsp
&\text{1 adjoint field} \to i^2 f^{qik} f^{kjq} = N \d^{ij}\\
&\text{$2N$ fundamental fields}\to 2N \Tr(T^iT^j) = N \d^{ij}.
\end{split}
\ee
Thus we are also free to decorate the diagrams of figure
\ref{fig:noloop} with one-loop-corrected propagators, see figure
\ref{fig:1loop}.

\begin{figure}[t]
\begin{center}
\includegraphics*[bb=25 50 325 190, height=0.9in]{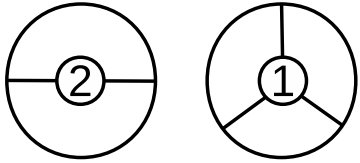}
\end{center}
\caption{After application of rules depicted in figures
  \ref{fig:noloop} and \ref{fig:1loop}, only the two-loop propagator,
  and 1-loop triple-vertex corrections remain at ${\cal O}(g^6)$.}
\label{fig:rest}
\end{figure}

It is worth underscoring at this point that we have now found
agreement with the matrix model results presented in section
\ref{sec:mm} at the first two consecutive orders of perturbation
theory, without evaluating a single Feynman diagram. At the next
order, ${\cal O}(g^6)$, we will have to do more work. Applying the
rules depicted in figures \ref{fig:noloop} and \ref{fig:1loop}, the
only diagrams remaining are bona fide two-loop matter\footnote{By
  ``matter'' we mean the fields in the hypermultiplet, whether they
  are in the adjoint or fundamental representation.} corrections to the
gauge/scalar propagator and bona fide one-loop matter corrections to the
triple vertex, see figure \ref{fig:rest}.
Let us concentrate on the former. We can reduce this class of diagram
even further. Introducing a fat graph notation, where fundamental
fields are represented by single lines, and adjoint ones by double
lines, we find that the following topology of diagram cancels between
the ${\cal N}=4$ and ${\cal N}=2$ theories

\begin{tabular}{rl}
\begin{minipage}{.5\textwidth}
\vspace{0.3cm}
\begin{center}
\includegraphics*[bb=50 45 490 150, height=0.5in]{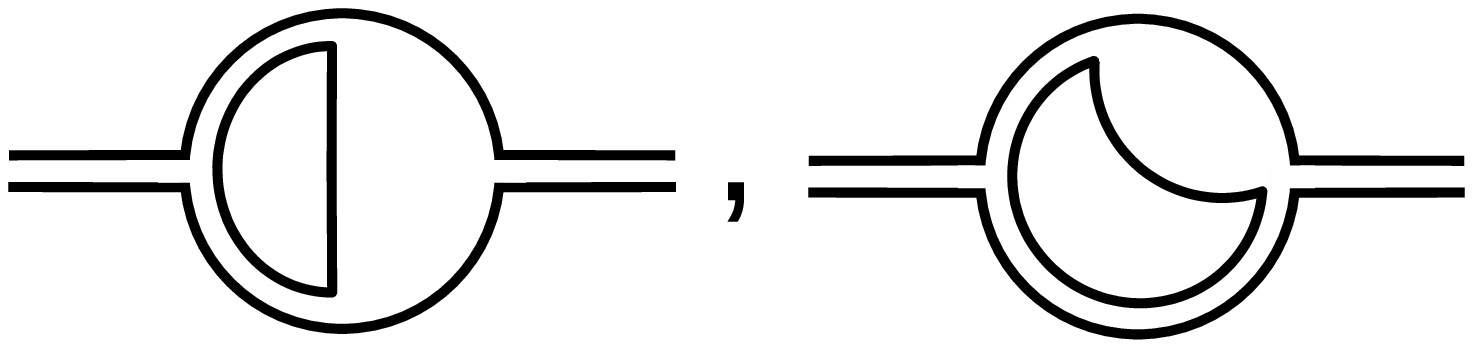}
\end{center}
\end{minipage}
\begin{minipage}{.4\textwidth}
\vspace{0.3cm}
\begin{center}
$\sim 2N\,i f^{qik} \Tr(T^k T^j T^q) = \frac{N^2}{2} \d^{ij}$ 
\end{center}
\end{minipage}
\end{tabular}\\

\ni whilst the adjoint counter-part has the same colour factor
\be
i^4 f^{qik} f^{klr} f^{rjm} f^{mlq} = \frac{N^2}{2} \d^{ij}.
\ee
For the two-loop matter corrections to the propagator, we find no
further cancellations. We are left with eight diagrams which are
collected and evaluated in appendix \ref{app:main}. 

It turns out that $\zeta(3)$ is very hard to come by in these Feynman
diagrams. In fact, the only time it appears is from the well-known
integral
\be\label{Iint}
{\cal I}=
\int \frac{d^4 k}{(2\pi)^4}\int \frac{d^4 q}{(2\pi)^4}
\frac{1}{k^2 q^2 (k-p)^2 (q-p)^2 (k-q)^2} 
= \frac{1}{(4\pi)^4p^2} \, 6 \,\zeta(3),
\ee
arising solely from the topology:
\begin{center}
\includegraphics*[bb=55 45 260 150, height=0.5in]{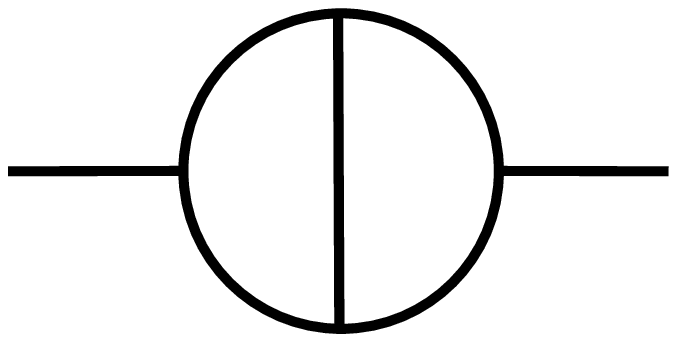}
\end{center}
which typically also contains other terms (owing to numerators),
however these other terms do not contain $\zeta(3)$. The general
structure of these diagrams is as follows
\be\label{gen}
\frac{(p^2)^{1-2\e}}{(4\pi)^{4-2\e}}
\left[ A_1 \left(\frac{1}{\e^2}-\zeta(2)\right) +\frac{A_2}{\e} +A_3 
+ A_4\,\zeta(3) +{\cal O}(\e)\right],
\ee
where $p$ is the external momentum, and where the $A_i$ are rational
numbers\footnote{The Euler-gamma terms have been removed through the
usual $e^{2\e\g}$ factor.}. This structure is also found for all the other
two-loop matter correction diagrams, albeit with $A_4=0$. In summing
the contributions from all diagrams we find that the coefficients
$A_1$, $A_2$, and $A_3$ sum to zero, and so all divergences and
non-$\zeta(3)$ terms cancel entirely. The details of the calculation
are collected in appendix \ref{app:main}.

The diagrams responsible for the $\zeta(3)$ terms are shown below,
where the solid (dashed) lines in the loop indicate the scalar
(fermion) fields of the hypermultiplet. The external lines represent
the adjoint gauge field (wiggly) or real scalar field
(straight)\footnote{Note that for fermion loops the real scalar field
  is also exchanged in the loop. For convenience we have let the
  vertical wiggly line represent both gauge and scalar exchange in
  this instance.\label{foot:sc}}. Let us begin by calculating the colour factor
associated with these diagrams. We are interested in the difference
between taking the matter in the adjoint and $2N$ times in the
fundamental, the result being
\bsp
&i^4 f^{qik} f^{klr} f^{rjm} f^{mlq} - 2N \Tr[T^i T^k T^j T^k]
=\frac{N^2}{2} \d^{ij} - 2N \frac{-\d^{ij}}{4N}=\frac{N^2+1}{2} \d^{ij}.
\end{split}
\ee 
Since in the perturbative expansion of the Wilson loop to second
order, i.e. two fields emanating from the loop, one gains a
factor of $\Tr(T^iT^i)/N \sim (N^2-1)/N$, one can already verify that
the correct colour factor has emerged for a match to
(\ref{mmres}). Explicitly we find that the colour-stripped diagram-differences
yield the following results
\begin{align*}
\parbox{32mm}{
\begin{fmfgraph*}(30,20)
\fmfleft{i}
\fmfright{o}
\fmf{dbl_wiggly, label=$\mu$, tension=2}{i,v1}
\fmf{dbl_wiggly, label=$\nu$, tension=2}{v3,o}
\fmf{dbl_plain_arrow,left=0.4, tension=1}{v1,v4,v3,v2,v1}
\fmffixed{(0,43)}{v2,v4}
\fmf{dbl_wiggly}{v2,v4}
\end{fmfgraph*}}-
\parbox{32mm}{
\begin{fmfgraph*}(30,20)
\fmfleft{i}
\fmfright{o}
\fmf{dbl_wiggly, label=$\mu$, tension=2}{i,v1}
\fmf{dbl_wiggly, label=$\nu$, tension=2}{v3,o}
\fmf{plain_arrow,left=0.4, tension=1}{v1,v4,v3,v2,v1}
\fmffixed{(0,43)}{v2,v4}
\fmf{dbl_wiggly}{v2,v4}
\end{fmfgraph*}} & = \frac{4}{3} \,p^4\,\left(\d^{\m\n}-\frac{p^\m
    p^\n}{p^2}\right)\, {\cal I}  +\text{non-$\zeta(3)$} \\
\parbox{32mm}{
\begin{fmfgraph*}(30,20)
\fmfleft{i}
\fmfright{o}
\fmf{dbl_wiggly, label=$\mu$, tension=2}{i,v1}
\fmf{dbl_wiggly, label=$\nu$, tension=2}{v3,o}
\fmf{dbl_dashes_arrow,left=0.4, tension=1}{v1,v4,v3,v2,v1}
\fmffixed{(0,43)}{v2,v4}
\fmf{dbl_wiggly}{v2,v4}
\end{fmfgraph*}}-
\parbox{32mm}{
\begin{fmfgraph*}(30,20)
\fmfleft{i}
\fmfright{o}
\fmf{dbl_wiggly, label=$\mu$, tension=2}{i,v1}
\fmf{dbl_wiggly, label=$\nu$, tension=2}{v3,o}
\fmf{dashes_arrow,left=0.4, tension=1}{v1,v4,v3,v2,v1}
\fmffixed{(0,43)}{v2,v4}
\fmf{dbl_wiggly}{v2,v4}
\end{fmfgraph*}} & = \frac{8}{3}\,p^4 \,\left(\d^{\m\n}-\frac{p^\m
    p^\n}{p^2}\right) \, {\cal I}  +\text{non-$\zeta(3)$}\\
\parbox{32mm}{
\begin{fmfgraph*}(30,20)
\fmfleft{i}
\fmfright{o}
\fmf{dbl_plain, label=$I$, tension=2}{i,v1}
\fmf{dbl_plain, label=$J$, tension=2}{v3,o}
\fmf{dbl_dashes_arrow,left=0.4, tension=1}{v1,v4,v3,v2,v1}
\fmffixed{(0,43)}{v2,v4}
\fmf{dbl_wiggly}{v2,v4}
\end{fmfgraph*}}-
\parbox{32mm}{
\begin{fmfgraph*}(30,20)
\fmfleft{i}
\fmfright{o}
\fmf{dbl_plain, label=$I$, tension=2}{i,v1}
\fmf{dbl_plain, label=$J$, tension=2}{v3,o}
\fmf{dashes_arrow,left=0.4, tension=1}{v1,v4,v3,v2,v1}
\fmffixed{(0,43)}{v2,v4}
\fmf{dbl_wiggly}{v2,v4}
\end{fmfgraph*}} & = 4\, p^4\, \d^{IJ}\, {\cal I}  +\text{non-$\zeta(3)$} \\
\end{align*}
where ${\cal I}$ is the expression given in (\ref{Iint}), and the
``non-$\zeta(3)$'' terms are the divergences and constant terms shown
schematically in (\ref{gen}), and which cancel identically against the
rest of the diagrams, which have no $\zeta(3)$ contribution. It is
clear that the gauge field and real scalar field propagator
contributions are equal, as they are in the celebrated one-loop
calculation for the ${\cal N}=4$ theory presented in
\cite{Erickson:2000af}. Adding the external propagators to these
amputated diagrams also reveals that the result is proportional to the
bare propagators\footnote{Up to the extra $p^\m p^\n/p^2$ factors,
  which may be removed by a gauge transformation.} $\d^{\m\n}/p^2$ and
$\d^{IJ}/p^2$, for the gauge, and real scalar fields respectively.
Fourier transforming back to position space and evaluating the
expectation value of the Wilson loop, one obtains
(\ref{mainres}). Plugging in the circular contour $x^\m =
(\cos\t,\sin\t,0,0)$, and $\Theta^I(\t) = \d^{I1}$, one obtains
%
%
%
\be\label{comb}
\frac{|\dot x_1||\dot x_2|  -  \dot x_1 \cdot\dot x_2}{4\pi^2
  (x_1-x_2)^2} = \frac{1}{4\pi^2} \frac{1}{2}.
\ee
The result is that (\ref{mainres}) is exactly the expression given in
(\ref{mmres}), namely
\be
\frac{3\, \zeta(3)}{512\pi^4} \frac{(N^2-1)(N^2+1)}{N},
\ee
and so we have recovered the matrix model result from perturbation
theory.

\subsection{One-loop corrected trivalent graph}
\label{sec:triv}

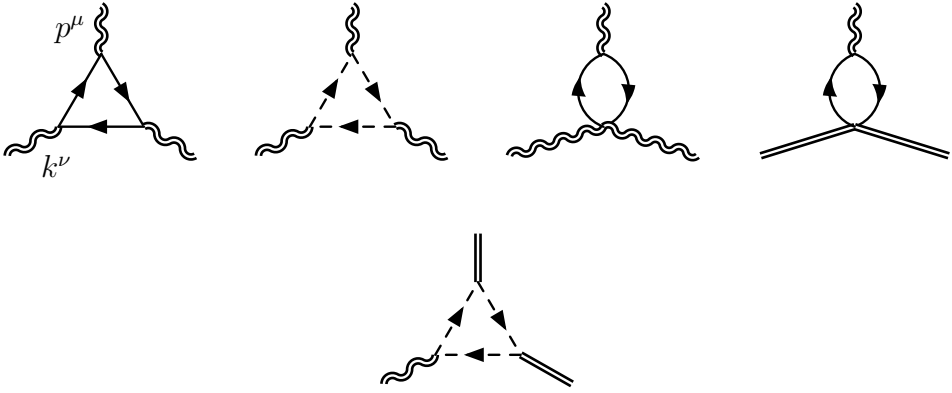
\begin{figure}[t]
\begin{center}
\parbox{32mm}{
\begin{fmfgraph*}(25,30)
\fmftop{t}
\fmfshift{(0,-.2w)}{t}
\fmfleft{i}
\fmfshift{(0,-.4w)}{i}
\fmfright{o}
\fmfshift{(0,-.4w)}{o}
\fmf{dbl_wiggly,label=$p^\m$, tension=1.2}{t,v1}
\fmf{plain_arrow, tension=0.4}{v1,v2}
\fmf{dbl_wiggly, tension=1}{v2,o}
\fmf{plain_arrow, tension=0.4}{v2,v3}
\fmf{dbl_wiggly,label=$k^\n$,tension=1}{i,v3}
\fmf{plain_arrow, tension=0.4}{v3,v1}
\end{fmfgraph*}}
\parbox{32mm}{
\begin{fmfgraph*}(25,30)
\fmftop{t}
\fmfshift{(0,-.2w)}{t}
\fmfleft{i}
\fmfshift{(0,-.4w)}{i}
\fmfright{o}
\fmfshift{(0,-.4w)}{o}
\fmf{dbl_wiggly, tension=1.2}{t,v1}
\fmf{dashes_arrow, tension=0.4}{v1,v2}
\fmf{dbl_wiggly, tension=1}{v2,o}
\fmf{dashes_arrow,  tension=0.4}{v2,v3}
\fmf{dbl_wiggly, tension=1}{i,v3}
\fmf{dashes_arrow, tension=0.4}{v3,v1}
\end{fmfgraph*}}
\parbox{32mm}{
\begin{fmfgraph*}(25,30)
\fmftop{t}
\fmfshift{(0,-.2w)}{t}
\fmfleft{i}
\fmfshift{(0,-.4w)}{i}
\fmfright{o}
\fmfshift{(0,-.4w)}{o}
\fmf{dbl_wiggly, tension=1.2}{t,v1}
\fmf{plain_arrow, left=0.7, tension=0.4}{v1,v2}
\fmf{dbl_wiggly, tension=1}{v2,o}
\fmf{dbl_wiggly, tension=1}{i,v2}
\fmf{plain_arrow, left=0.7, tension=0.4}{v2,v1}
\end{fmfgraph*}}
\parbox{32mm}{
\begin{fmfgraph*}(25,30)
\fmftop{t}
\fmfshift{(0,-.2w)}{t}
\fmfleft{i}
\fmfshift{(0,-.4w)}{i}
\fmfright{o}
\fmfshift{(0,-.4w)}{o}
\fmf{dbl_wiggly, tension=1.2}{t,v1}
\fmf{plain_arrow, left=0.7, tension=0.4}{v1,v2}
\fmf{dbl_plain, tension=1}{v2,o}
\fmf{dbl_plain, tension=1}{i,v2}
\fmf{plain_arrow, left=0.7, tension=0.4}{v2,v1}
\end{fmfgraph*}}
\parbox{32mm}{
\begin{fmfgraph*}(25,30)
 \fmfset{arrow_len}{8}\fmfset{arrow_ang}{20}
\fmftop{t}
\fmfshift{(0,-.2w)}{t}
\fmfleft{i}
\fmfshift{(0,-.4w)}{i}
\fmfright{o}
\fmfshift{(0,-.4w)}{o}
\fmf{dbl_plain, tension=1.2}{t,v1}
\fmf{dashes_arrow, tension=0.4}{v1,v2}
\fmf{dbl_plain, tension=1}{v2,o}
\fmf{dashes_arrow, tension=0.4}{v2,v3}
\fmf{dbl_wiggly, tension=1}{i,v3}
\fmf{dashes_arrow, tension=0.4}{v3,v1}
\end{fmfgraph*}}
\end{center}
\caption{One-matter-loop corrections to the triple vertex, shown
  (to reduce clutter) for the ${\cal N}=2$ theory only. In the
  ${\cal N}=4$ case internal lines are doubled.}
\label{fig:triv}
\end{figure}

We now turn our attention to the one-matter-loop corrected trivalent
graph, contributions to which are shown in figure \ref{fig:triv}.
This diagram
presents an interesting manifestation of the difference between
$SU(2)$ and $SU(N)$ for $N>2$. Let us look at the colour factors
arising from the ${\cal N}=4$ and ${\cal N}=2$ versions of this
graph. We have
\bsp
&{\cal N}=4 ~\to~ i^3 f^{qim} f^{mjn} f^{nkq} = \frac{N}{2} i f^{ijk},\\
&{\cal N}=2 ~\to~ 2N \Tr(T^i T^j T^k) = \frac{N}{2} i f^{ijk} +
\frac{N}{2} d^{ijk}.\\
\end{split}
\ee
We see immediately that for $SU(2)$, where $d^{ijk}=0$, these diagrams
cancel identically. But for $SU(N)$ with $N>2$, we are left with a
term proportional to the totally symmetric structure constant
$d^{ijk}$. This immediately implies that the result must also be
symmetric in interchange between the two momenta $p^\m$ and
$k^\n$, see figure \ref{fig:triv}. In the Wilson loop expanded to 3$^{\text{rd}}$ order, we will encounter
another $\Tr(T^m T^n T^q) \sim if^{mnq} + d^{mnq}$. Clearly only the
$d^{mnq}$ can survive, as the trace will be contracted with the $d^{ijk}$
coming from the loop-corrected vertex. This leaves us with a
completely symmetrized sum of path orderings in the Wilson loop
expansion, which means that the path-ordering is removed, and we have
complete integrals over each of the insertion points. The situation is
most easily seen for the triple vertex with two real scalar fields
and one gauge field\footnote{Generalizing to the case of three external gauge fields does
not alter the result.}. It is clear that the corrected vertex must be of
the form
\begin{align*}
\parbox{32mm}{
\begin{fmfgraph*}(25,30)
\fmftop{t}
\fmfshift{(0,-.2w)}{t}
\fmfleft{i}
\fmfshift{(0,-.2w)}{i}
\fmfright{o}
\fmfshift{(0,-.2w)}{o}
\fmf{dbl_plain, label=$I$, tension=1}{t,v1}
\fmf{dbl_plain, label=$J$, tension=1}{v1,o}
\fmf{dbl_wiggly, label=$\mu$, tension=1}{i,v1}
\fmfv{d.sh=circle,l.d=0, d.f=empty,d.si=.4w,l=\tiny{${\cal N}=4$}}{v1}
\end{fmfgraph*}}-
\parbox{32mm}{
\begin{fmfgraph*}(25,30)
\fmftop{t}
\fmfshift{(0,-.2w)}{t}
\fmfleft{i}
\fmfshift{(0,-.2w)}{i}
\fmfright{o}
\fmfshift{(0,-.2w)}{o}
\fmf{dbl_plain, label=$I$, tension=1}{t,v1}
\fmf{dbl_plain, label=$J$, tension=1}{v1,o}
\fmf{dbl_wiggly, label=$\mu$, tension=1}{i,v1}
\fmfv{d.sh=circle,l.d=0, d.f=empty,d.si=.4w,l=\tiny{${\cal N}=2$}}{v1}
\end{fmfgraph*}}
 & =  \d^{IJ}\,(p^\m + k^\m)\, F(p,k)
\end{align*}
where $F(p,k) = F(k,p)$. Decorating with propagators and Fourier
transforming to position space, we will have, schematically
\be
~ \p_{x_3^\mu} \int d^4p\,d^4k\, e^{ip\cdot(x_1-x_3) + ik \cdot (x_2-x_3)}
\,\frac{F(p,k)}{p^2 k^2 (p+k)^2}.
\ee
But then this will be integrated over in the Wilson loop
\be
\oint d\t_3 \, \dot x_3 \cdot \p_{x_3}  \int d^4p\,d^4k\, e^{ip\cdot(x_1-x_3) + ik \cdot (x_2-x_3)}
\,\frac{F(p,k)}{p^2 k^2 (p+k)^2} = 0
\ee
which is the integral of a total-derivative and therefore
vanishes. Note that this cancellation is strictly only
true for a closed loop, an open loop could give boundary
contributions.

\section{Comments on light-like loops and loops
on $\bR^2$ and $S^2$}
\label{sec:scamp}

As stressed in the introduction, the difference between the Wilson
loop in ${\cal N}=4$ and superconformal ${\cal N}=2$ SYM appears as a
term proportional to the bare gauge field (and associated real scalar)
propagator. This fact is independent of the shape of the Wilson loop,
although we were primarily interested in the circle for obvious
reasons. In this section we would like to point-out a couple of
implications of this result.

\subsection{Zarembo loops in the plane}

The first is that we could consider Wilson loops of various shape, as
long as the scalar coupling $\Theta^I(\t)$ remains on an $S^1$. This
is because in the ${\cal N}=2$ theory, we only have two real scalar
fields $\Phi_I$. This class includes the planar Zarembo loops
\cite{Zarembo:2002an,Agarwal:2009up} of arbitrary shape, and the
longitudes Wilson loop of \cite{Drukker:2007qr}. The Zarembo
loops are defined by
\be
|\dot x|\, \Theta^I(\t) = M_\m^I \,\dot x^\m ,\quad
M_\m^I M^I_\n = \d_{\m\n},
\ee
and so have the property that the combined gauge field and scalar
exchange, i.e. the LHS of (\ref{comb}), is identically zero. This
leads to the result $\la W_{\text{Zarembo}}\ra=1$, which is true to
all orders in perturbation theory
\cite{Guralnik:2003di,Guralnik:2004yc}. It is clear that the
triviality of the expectation value will not be disturbed in the
superconformal ${\cal N}=2$ case at ${\cal O}(g^6)$, since the
corrected scalar and gauge field propagators remain equal. This makes
sense, as the supersymmetry respected by this Wilson loop in the
${\cal N}=2$ theory is the same as that in the ${\cal N}=4$ theory.

\subsection{Longitudes Wilson loop}

The longitudes Wilson loop is given by an ``orange wedge'', descending
from the north pole of an $S^2$ along a great circle to the south
pole, and then returning along a second longitude shifted by an
azimuthal angle $\a$
\bsp
&x^\m = \begin{cases}
(\sin\t,0,\cos\t,0),\qquad &0 \leq \t \leq \pi,\\
(-\cos\a\sin\t,-\sin\a\sin\t,\cos\t,0),\qquad &\pi \leq \t \leq 2\pi.
\end{cases}\\
&|\dot x| \Theta^I \Phi_I = \begin{cases}
\Phi_2,\qquad &0 \leq \t \leq \pi,\\
-\Phi_2\cos\a+\Phi_1\sin\a,\qquad &\pi \leq \t \leq 2\pi.
\end{cases}\\
\end{split}
\ee
As conjectured in \cite{Drukker:2007dw,Drukker:2007yx,Drukker:2007qr}, and
backed-up in
\cite{Bassetto:2008yf,Young:2008ed,Bassetto:2009rt,Bassetto:2009ms,Giombi:2009ms,Giombi:2009ds,Pestun:2009nn},
it seems very certain that these Wilson loops in ${\cal N}=4$ SYM are
captured completely by pure two-dimensional Yang-Mills theory, and
therefore enjoy invariance under area-preserving diffeomorphisms. This
results in an expectation value which depends only on the area enclosed
by the longitudes. Therefore the ${\cal O}(g^2)$ term is proportional
to this area, and so then is the correction introduced by the ${\cal
  N}=2$ superconformal theory at ${\cal O}(g^6)$. This observation
leads one to the possibility that there may exist some deformation of
pure 2-d Yang-Mills or of the correspondence between it and the Wilson
loops of \cite{Drukker:2007qr} which would accommodate the ${\cal N}=2$
superconformal analogue, perhaps as a 1-loop determinant factor to be
introduced into the localized path integral of \cite{Pestun:2009nn}. 

\subsection{Light-like Wilson loops and scattering amplitudes}

As a final remark it is interesting to consider light-like polygonal
Wilson loops in the superconformal ${\cal N}=2$ theory, and any
possible connection they may have to scattering amplitudes. We have
proven that at ${\cal O}(g^4)$ there is no difference between the
${\cal N}=4$ and ${\cal N}=2$ results for Wilson loops. However, in
the ${\cal N}=4$ theory, and in the planar limit, we know that a
light-like Wilson loop at ${\cal O}(g^4)$ is equivalent to a gluon
scattering amplitude at two-loops \cite{Drummond:2007cf}. It is
interesting to ask whether gluon scattering in the ${\cal N}=2$
superconformal theory at large-$N$, i.e. in the Veneziano limit, is
modified with respect to ${\cal N}=4$ SYM, and more importantly at
which order in perturbation theory. In the work \cite{Glover:2008tu},
it was shown that there is no modification at one-loop, see
\cite{Lal:2009gn} for related work. Based on our considerations of
section \ref{sec:pt}, at one-loop, we showed that propagator loop
corrections cancel, but the corrected triple vertex does not
necessarily cancel for $N>2$. It would be interesting to understand
how this becomes consistent with \cite{Glover:2008tu}.

In order for there to be an analogous scattering amplitude/Wilson loop
duality for the ${\cal N}=2$ theory at large-$N$, the gluon scattering
amplitudes would have to be the same in the two theories at two-loops,
and different (by exactly the term given in (\ref{mainres}), evaluated
for a light-like polygonal contour) at three-loops. It would be
necessary to have higher loop versions of the results in
\cite{Glover:2008tu} in order to check this. In any case, it is
interesting to continue to investigate gluon scattering amplitudes in
the ${\cal N}=2$ theory. There are indications that the theory may be
integrable in the planar limit \cite{Gadde:2010zi}, and there is also
work to identify a string dual \cite{Gadde:2009dj}.

\section*{Acknowledgements}

We would like to thank Jan Plefka, Johannes Henn, and Filippo
Passerini for discussions. This work is supported by the Volkswagen
Foundation.


\appendix

\section{Two-loop matter corrections}
\label{app:main}

We write the action of Euclidean ${\cal N}=4$ SYM following \cite{Rey:2010ry},
as the sum of an ${\cal N}=1$ SYM in six dimensions (consisting of a
6-d gauge field $A_M$ and an 8-component Majorana-Weyl spinor $\l$)
dimensionally reduced to $d=4-2\e$, and a single adjoint
hypermultiplet consisting of 2 complex scalar fields $q^\a$, $\a=1,2$,
$(q^\a)^\dag=q_\a$, and a complex four-dimensional (four-component)
spinor $\Psi$
\bsp\label{n4act}
S_{{\cal N}=4} = \frac{2}{g^2} &\Tr \int d^d x \, \Biggl[
\frac{1}{4} F_{MN}^2 + \frac{i}{2} \bar \l \G^M D_M \l
+ D_M q^\a D_M q_\a + i \bar \Psi \G^M D_M \Psi\\
& + \bar \l \g^\a[q_\a,\Psi] + \bar\Psi \g_\a [q^\a,\l]
+[q_\a,q^\b][q_\b,q^\a]- \frac{1}{2}[q_\a,q^\a][q_\b,q^\b]
\Biggr],
\end{split}
\ee 
where the $\g^\a$ are a set of gamma matrices obeying
$\{\g^\a,\g_\b\}=2\d^\a_\b$ and anti-commuting with all $\G^M$, see
\cite{Rey:2010ry} for details. To deform this theory to the ${\cal
  N}=2$ superconformal SYM we take $2N$ hypermultiplets instead of
one, so that the $q$ and $\Psi$ fields earn a flavour index and become
vectors, and we take them in the fundamental, instead of the adjoint
representation
\bsp\label{n2act}
S_{{\cal N}=2} = \frac{1}{g^2} \int d^d x \, \Biggl[&
2\Tr\left(\frac{1}{4} F_{MN}^2 + \frac{i}{2} \bar \l \G^M D_M \l\right)
+ D_M \vec q^\a \cdot D_M \vec q_\a + i \vec{\bar \Psi} \cdot \G^M D_M \vec \Psi\\
& - \bar \l^i \g^\a \vec q_\a \cdot T^i \vec\Psi -\vec{\bar \Psi}
\cdot \g_\a T^i \vec q^\a \l^i\\
&+\left(\vec q_\a \cdot T^i \vec q^\b\right)\left( \vec q_\b \cdot T^i \vec q^\a\right)
-\frac{1}{2}\left(\vec q_\a \cdot T^i \vec q^\a\right)\left( \vec q_\b \cdot T^i \vec q^\b\right)
\Biggr],
\end{split}
\ee 
where we have introduced the standard $SU(N)$ generators $T^i$,
$i=1,\ldots,N^2-1$, obeying
\bsp
&T^i T^i = \frac{N^2-1}{2N} {\bf 1}, \quad \Tr(T^iT^j) = \frac{1}{2}
\d^{ij}, \quad [T^i,T^j] = i f^{ijk} T^k, \quad f^{ijk}f^{ijl} =
N\d^{kl},\\
&\{T^i,T^j\} = \frac{1}{N}\d^{ij} {\bf 1} + d^{ijk} T^k ,
\end{split}
\ee
and the covariant derivatives act as follows
\bsp
&D_M \vec q^\a = \p_M \vec q^\a -i A_M^i T^i \vec q^\a,\quad
D_M \vec q_\a = \p_M \vec q_\a +i \vec q_\a  T^i A_M^i, \\
&D_M \vec \Psi = \p_M \vec \Psi - i A^i_M T^i \vec \Psi.
\end{split}
\ee
The pure gauge portion of the two actions (\ref{n4act}) and
(\ref{n2act}) is exactly the same. This means that in the diagrammatic
difference, ghosts cancel trivially. We are free to work in Feynman
gauge where the propagators are
\bsp
&\la q^\a q_\b \ra = g^2 \frac{\d^\a_\b}{p^2},\quad
\la \Psi \bar\Psi \ra = -g^2 \frac{\G^\m p_\m}{p^2},\\
&\la A^i_M A^j_N \ra = g^2 \frac{\d^{ij} \d_{MN}}{p^2},
\quad \la \l^i \bar\l^j \ra = -g^2 \d^{ij}\frac{\G^\m p_\m}{p^2},
\end{split}
\ee 
and where there is an implied delta function on the hypermultiplet
propagators for either adjoint, or fundamental and flavour indices.

\subsection{The diagrams}

There are 8 diagrams which do not cancel trivially between the ${\cal
  N}=4$ and ${\cal N}=2$ theories. In this section we list the results
for each of them. Every diagram gives a common factor of
\be
\frac{(p^2)^{1-2\e}}{(4\pi)^d} (N^2+1) \d^{ij},
\ee 
where $\d^{ij}$ is the delta function on the colour indices. We
suppress this factor below. The results are given in terms of the
scale-free integrals defined by \cite{Grozin:2005yg}
\be
G(n_1,n_2,n_3,n_4,n_5) \equiv \frac{1}{\pi^d}\int \frac{d^d k\, d^d q}
{(k^2)^{n_1} (q^2)^{n_2} ((k-p)^2)^{n_3} ((q-p)^2)^{n_4} ((q-k)^2)^{n_5}}
\ee
where in the above expression we replace $p^2 \to 1$. These integrals
may be reduced to products of one-loop integrals using well-known
techniques \cite{Grozin:2005yg}\footnote{The integral from which the
  $\zeta(3)$ comes, i.e. (\ref{Iint}), is $G(1,1,1,1,1)$.}. In the case of corrections to the
gauge field propagator, which we present first, we take the trace over
the external space-time indices $\m$, and $\n$. We have verified that
the projection onto $p_\m$, $p_\n$ yields the same cancellation of
non-$\zeta(3)$ terms, and yields no $\zeta(3)$ contribution. Note that
the comments of footnote \ref{foot:sc} apply equally to the diagrams
below.

The external lines are either wiggly (gauge field $A_\m$) or straight
(scalar field $\Phi_I$), while the internal lines carry arrows and are
straight for the hypermultiplet scalar $q^\a$ and dashed for the
hypermultiplet fermion $\Psi$. The dotted line denotes the gaugino
$\l$. We take the ``diagrammatic difference'' between the ${\cal N}=4$
and ${\cal N}=2$ theories, and thus report the differences between the
diagrams. The fundamental representation is indicated by single, as
opposed to double, lines.  

\begin{align*}
&\parbox{32mm}{
\begin{fmfgraph*}(30,20)
\fmfleft{i}
\fmfright{o}
\fmf{dbl_wiggly, label=$\mu$, tension=2}{i,v1}
\fmf{dbl_wiggly, label=$\mu$, tension=2}{v3,o}
\fmf{dbl_plain_arrow,left=0.4, tension=1}{v1,v4,v3,v2,v1}
\fmffixed{(0,43)}{v2,v4}
\fmf{dbl_wiggly}{v2,v4}
\end{fmfgraph*}}-
\parbox{32mm}{
\begin{fmfgraph*}(30,20)
\fmfleft{i}
\fmfright{o}
\fmf{dbl_wiggly, label=$\mu$, tension=2}{i,v1}
\fmf{dbl_wiggly, label=$\mu$, tension=2}{v3,o}
\fmf{plain_arrow,left=0.4, tension=1}{v1,v4,v3,v2,v1}
\fmffixed{(0,43)}{v2,v4}
\fmf{dbl_wiggly}{v2,v4}
\end{fmfgraph*}}  = 4 G(-1,1,1,1,1)+4 G(0,1,1,0,1)\\
&-12 G(0,1,1,1,1)+2 G(1,1,1,1,-1)+5 G(1,1,1,1,0)+2 G(1,1,1,1,1)\\
&\parbox{32mm}{
\begin{fmfgraph*}(30,20)
\fmfleft{i}
\fmfright{o}
\fmf{dbl_wiggly, label=$\mu$, tension=2}{i,v1}
\fmf{dbl_wiggly, label=$\mu$, tension=2}{v3,o}
\fmf{dbl_dashes_arrow,left=0.4, tension=1}{v1,v4,v3,v2,v1}
\fmffixed{(0,43)}{v2,v4}
\fmf{dbl_wiggly}{v2,v4}
\end{fmfgraph*}}-
\parbox{32mm}{
\begin{fmfgraph*}(30,20)
\fmfleft{i}
\fmfright{o}
\fmf{dbl_wiggly, label=$\mu$, tension=2}{i,v1}
\fmf{dbl_wiggly, label=$\mu$, tension=2}{v3,o}
\fmf{dashes_arrow,left=0.4, tension=1}{v1,v4,v3,v2,v1}
\fmffixed{(0,43)}{v2,v4}
\fmf{dbl_wiggly}{v2,v4}
\end{fmfgraph*}}  = 4 [-2 (\e-1) G(0,1,1,0,1)+4 (\e-1) G(0,1,1,1,1)\\
&-(\e-1) G(1,1,1,1,1)+2 [G(1,1,1,1,-1)+G(1,1,1,1,0)]]\\
&\parbox{32mm}{
\begin{fmfgraph*}(30,20)
 \fmfset{arrow_len}{8}\fmfset{arrow_ang}{20}
\fmfleft{i}
\fmfright{o}
\fmf{dbl_wiggly, label=$\mu$, tension=2}{i,v1}
\fmf{dbl_wiggly, label=$\mu$, tension=2}{v3,o}
\fmf{dbl_dashes_arrow,left=0.4, tension=1}{v4,v3,v2}
\fmf{dbl_plain_arrow,left=0.4, tension=1}{v2,v1,v4}
\fmffixed{(0,43)}{v2,v4}
\fmf{dbl_dots}{v4,v2}
\end{fmfgraph*}} - 
\parbox{32mm}{
\begin{fmfgraph*}(30,20)
 \fmfset{arrow_len}{8}\fmfset{arrow_ang}{20}
\fmfleft{i}
\fmfright{o}
\fmf{dbl_wiggly, label=$\mu$, tension=2}{i,v1}
\fmf{dbl_wiggly, label=$\mu$, tension=2}{v3,o}
\fmf{dashes_arrow,left=0.4, tension=1}{v4,v3,v2}
\fmf{plain_arrow,left=0.4, tension=1}{v2,v1,v4}
\fmffixed{(0,43)}{v2,v4}
\fmf{dbl_dots}{v4,v2}
\end{fmfgraph*}}
=16 (G(0,1,1,0,1)-G(1,1,1,1,-1)-G(1,1,1,1,0))\\
\end{align*}
\begin{align*}
&\parbox{32mm}{
\begin{fmfgraph*}(30,20)
\fmfleft{i}
\fmfright{o}
\fmf{dbl_wiggly,label=$\mu$,tension=5}{i,v1}
\fmf{dbl_wiggly,label=$\mu$,tension=5}{v2,o}
\fmf{dbl_plain_arrow,left,tension=0.7}{v1,v2,v1}
\fmf{dbl_wiggly}{v1,v2}
\end{fmfgraph*}} - 
\parbox{32mm}{
\begin{fmfgraph*}(30,20)
\fmfleft{i}
\fmfright{o}
\fmf{dbl_wiggly, label=$\mu$,tension=5}{i,v1}
\fmf{dbl_wiggly, label=$\mu$,tension=5}{v2,o}
\fmf{plain_arrow,left,tension=0.7}{v1,v2,v1}
\fmf{dbl_wiggly}{v1,v2}
\end{fmfgraph*}}
=-8 (\e-2) G(0,1,1,0,1)\\
&\parbox{32mm}{
\begin{fmfgraph*}(30,20)
\fmfleft{i}
\fmfright{o}
\fmftop{t}
\fmf{dbl_wiggly, label=$\m$, tension=2}{i,v1}
\fmf{dbl_wiggly, label=$\m$, tension=2}{v3,o}
\fmf{dbl_plain_arrow,left=1, tension=1}{v3,v1}
\fmffreeze
\fmf{dbl_plain_arrow, left=0.39, tension=0.12}{v1,v2}
\fmf{dbl_plain_arrow, left=0.39, tension=0.12}{v2,v3}
\fmf{phantom,tension=0.8}{t,v2}
\fmf{dbl_wiggly, right=0.6, tension=0.0}{v1,v2}
\end{fmfgraph*}} -
\parbox{32mm}{
\begin{fmfgraph*}(30,20)
\fmfleft{i}
\fmfright{o}
\fmftop{t}
\fmf{dbl_wiggly, label=$\m$, tension=2}{i,v1}
\fmf{dbl_wiggly, label=$\m$, tension=2}{v3,o}
\fmf{plain_arrow,left=1, tension=1}{v3,v1}
\fmffreeze
\fmf{plain_arrow, left=0.39, tension=0.12}{v1,v2}
\fmf{plain_arrow, left=0.39, tension=0.12}{v2,v3}
\fmf{phantom,tension=0.8}{t,v2}
\fmf{dbl_wiggly, right=0.6, tension=0.0}{v1,v2}
\end{fmfgraph*}}
=-4 (G(-1,1,1,1,1)+5 G(0,1,1,0,1)-2 G(0,1,1,1,1))\\
&\parbox{32mm}{
\begin{fmfgraph*}(30,20)
 \fmfset{arrow_len}{8}\fmfset{arrow_ang}{20}
\fmfleft{i}
\fmfright{o}
\fmf{dbl_wiggly, label=$\m$,tension=2}{i,v1}
\fmf{dbl_wiggly, label=$\m$,tension=2}{v3,o}
\fmf{dbl_plain_arrow,left,tension=1}{v1,v2,v1}
\fmf{dbl_plain_arrow,left,tension=1}{v2,v3,v2}
\end{fmfgraph*}} -
\parbox{32mm}{
\begin{fmfgraph*}(30,20)
 \fmfset{arrow_len}{8}\fmfset{arrow_ang}{20}
\fmfleft{i}
\fmfright{o}
\fmf{dbl_wiggly, label=$\m$,tension=2}{i,v1}
\fmf{dbl_wiggly, label=$\m$,tension=2}{v3,o}
\fmf{plain_arrow,left,tension=1}{v1,v2,v1}
\fmf{plain_arrow,left,tension=1}{v2,v3,v2}
\end{fmfgraph*}}
=3 (2 G(1,1,1,1,-1)+G(1,1,1,1,0))\\
\end{align*}
\begin{align*}
&\parbox{32mm}{
\begin{fmfgraph*}(30,20)
\fmfleft{i}
\fmfright{o}
\fmftop{t}
\fmf{dbl_wiggly, label=$\mu$ ,tension=1.4}{i,v1}
\fmf{dbl_wiggly,label=$\mu$,tension=1.4}{v2,o}
\fmf{dbl_plain_arrow, right,tension=1}{v1,v2}
\fmffreeze
\fmf{phantom, tension=.8}{t,v3}
\fmf{dbl_plain_arrow, right=0.4,tension=0.2}{v3,v1}
\fmf{dbl_plain_arrow, right=.4,tension=0.2}{v2,v3}
\fmfv{d.sh=circle,l.d=0, d.f=empty,d.si=.2w,l=$1$}{v3}
\end{fmfgraph*}} -
\parbox{32mm}{
\begin{fmfgraph*}(30,20)
\fmfleft{i}
\fmfright{o}
\fmftop{t}
\fmf{dbl_wiggly, label=$\mu$ ,tension=1.4}{i,v1}
\fmf{dbl_wiggly,label=$\mu$,tension=1.4}{v2,o}
\fmf{plain_arrow, right,tension=1}{v1,v2}
\fmffreeze
\fmf{phantom, tension=.8}{t,v3}
\fmf{plain_arrow, right=0.4,tension=0.2}{v3,v1}
\fmf{plain_arrow, right=.4,tension=0.2}{v2,v3}
\fmfv{d.sh=circle,l.d=0, d.f=empty,d.si=.2w,l=$1$}{v3}
\end{fmfgraph*}} 
=4 G(0,1,1,1,1)-8 G(0,1,1,0,1)\\
&\parbox{32mm}{
\begin{fmfgraph*}(30,20)
\fmfleft{i}
\fmfright{o}
\fmftop{t}
\fmf{dbl_wiggly, label=$\mu$ ,tension=1.4}{i,v1}
\fmf{dbl_wiggly,label=$\mu$,tension=1.4}{v2,o}
\fmf{dbl_dashes_arrow, right,tension=1}{v1,v2}
\fmffreeze
\fmf{phantom, tension=.8}{t,v3}
\fmf{dbl_dashes_arrow, right=0.4,tension=0.2}{v3,v1}
\fmf{dbl_dashes_arrow, right=.4,tension=0.2}{v2,v3}
\fmfv{d.sh=circle,l.d=0, d.f=empty,d.si=.2w,l=$1$}{v3}
\end{fmfgraph*}} -
\parbox{32mm}{
\begin{fmfgraph*}(30,20)
\fmfleft{i}
\fmfright{o}
\fmftop{t}
\fmf{dbl_wiggly, label=$\mu$ ,tension=1.4}{i,v1}
\fmf{dbl_wiggly,label=$\mu$,tension=1.4}{v2,o}
\fmf{dashes_arrow, right,tension=1}{v1,v2}
\fmffreeze
\fmf{phantom, tension=.8}{t,v3}
\fmf{dashes_arrow, right=0.4,tension=0.2}{v3,v1}
\fmf{dashes_arrow, right=.4,tension=0.2}{v2,v3}
\fmfv{d.sh=circle,l.d=0, d.f=empty,d.si=.2w,l=$1$}{v3}
\end{fmfgraph*}} 
=32 (\e-1) [G(-1,1,1,1,1)+G(0,1,1,0,1)\\
&\qquad\qquad\qquad\qquad\qquad\qquad\qquad\qquad\qquad
-G(0,1,1,1,1)]
\end{align*}
\begin{align*}
&\parbox{32mm}{
\begin{fmfgraph*}(30,20)
\fmfleft{i}
\fmfright{o}
\fmf{dbl_plain, label=$I$, tension=2}{i,v1}
\fmf{dbl_plain, label=$J$, tension=2}{v3,o}
\fmf{dbl_dashes_arrow,left=0.4, tension=1}{v1,v4,v3,v2,v1}
\fmffixed{(0,43)}{v2,v4}
\fmf{dbl_wiggly}{v2,v4}
\end{fmfgraph*}}-
\parbox{32mm}{
\begin{fmfgraph*}(30,20)
\fmfleft{i}
\fmfright{o}
\fmf{dbl_plain, label=$I$, tension=2}{i,v1}
\fmf{dbl_plain, label=$J$, tension=2}{v3,o}
\fmf{dashes_arrow,left=0.4, tension=1}{v1,v4,v3,v2,v1}
\fmffixed{(0,43)}{v2,v4}
\fmf{dbl_wiggly}{v2,v4}
\end{fmfgraph*}} 
=2\d^{IJ} (2 G(0,1,1,0,1)-4 G(0,1,1,1,1)+G(1,1,1,1,1))\\
&\parbox{32mm}{
\begin{fmfgraph*}(30,20)
\fmfleft{i}
\fmfright{o}
\fmf{dbl_plain,label=$I$,tension=5}{i,v1}
\fmf{dbl_plain,label=$J$,tension=5}{v2,o}
\fmf{dbl_plain_arrow,left,tension=0.7}{v1,v2,v1}
\fmf{dbl_plain}{v1,v2}
\end{fmfgraph*}} - 
\parbox{32mm}{
\begin{fmfgraph*}(30,20)
\fmfleft{i}
\fmfright{o}
\fmf{dbl_plain, label=$I$,tension=5}{i,v1}
\fmf{dbl_plain, label=$J$,tension=5}{v2,o}
\fmf{plain_arrow,left,tension=0.7}{v1,v2,v1}
\fmf{dbl_plain}{v1,v2}
\end{fmfgraph*}}
=4\d^{IJ} G(0,1,1,0,1)\\
&\parbox{32mm}{
\begin{fmfgraph*}(30,20)
\fmfleft{i}
\fmfright{o}
\fmftop{t}
\fmf{dbl_plain, label=$I$ ,tension=1.4}{i,v1}
\fmf{dbl_plain,label=$J$,tension=1.4}{v2,o}
\fmf{dbl_dashes_arrow, right,tension=1}{v1,v2}
\fmffreeze
\fmf{phantom, tension=.8}{t,v3}
\fmf{dbl_dashes_arrow, right=0.4,tension=0.2}{v3,v1}
\fmf{dbl_dashes_arrow, right=.4,tension=0.2}{v2,v3}
\fmfv{d.sh=circle,l.d=0, d.f=empty,d.si=.2w,l=$1$}{v3}
\end{fmfgraph*}} -
\parbox{32mm}{
\begin{fmfgraph*}(30,20)
\fmfleft{i}
\fmfright{o}
\fmftop{t}
\fmf{dbl_plain, label=$I$ ,tension=1.4}{i,v1}
\fmf{dbl_plain,label=$J$,tension=1.4}{v2,o}
\fmf{dashes_arrow, right,tension=1}{v1,v2}
\fmffreeze
\fmf{phantom, tension=.8}{t,v3}
\fmf{dashes_arrow, right=0.4,tension=0.2}{v3,v1}
\fmf{dashes_arrow, right=.4,tension=0.2}{v2,v3}
\fmfv{d.sh=circle,l.d=0, d.f=empty,d.si=.2w,l=$1$}{v3}
\end{fmfgraph*}} 
=-16 \d^{IJ} (G(-1,1,1,1,1)+G(0,1,1,0,1)\\
&\qquad\qquad\qquad\qquad\qquad\qquad\qquad\qquad\qquad\qquad
-G(0,1,1,1,1))
\end{align*}
In summing these contributions we find that all non-$\zeta(3)$
contributions, including divergences, canel identically.  The 1-loop
corrections to the hypermultiplet propagators indicated above, are
given explicitly by the following diagrams (we show only the ${\cal
  N}=4$ case)
\begin{align*}
\parbox{32mm}{
\vspace{-1.20cm}
\begin{fmfgraph*}(30,20)
\fmfleft{i}
\fmfshift{(0,-.2w)}{i}
\fmfright{o}
\fmfshift{(0,-.2w)}{o}
\fmf{dbl_plain_arrow, tension=1}{v1,o}
\fmf{dbl_plain_arrow, tension=1}{i,v1}
\fmfv{d.sh=circle,l.d=0, d.f=empty,d.si=.3w,l=$1$}{v1}
\end{fmfgraph*}}=
\parbox{32mm}{
\begin{fmfgraph*}(30,20)
 \fmfset{arrow_len}{8}\fmfset{arrow_ang}{20}
\fmfleft{i}
\fmfright{o}
\fmf{dbl_plain_arrow ,tension=3}{i,v1}
\fmf{dbl_plain_arrow ,tension=3}{v2,o}
\fmf{dbl_wiggly,left,tension=1}{v1,v2}
\fmf{dbl_plain_arrow,tension=1}{v1,v2}
\end{fmfgraph*}} +
\parbox{32mm}{
\begin{fmfgraph*}(30,20)
\fmfleft{i}
\fmfright{o}
\fmf{dbl_plain_arrow,tension=3}{i,v1}
\fmf{dbl_plain_arrow,tension=3}{v2,o}
\fmf{dbl_dashes_arrow,left,tension=1}{v1,v2}
\fmf{dbl_dots,left,tension=1}{v2,v1}
\end{fmfgraph*}},\\
\parbox{32mm}{
\vspace{-1.20cm}
\begin{fmfgraph*}(30,20)
\fmfleft{i}
\fmfshift{(0,-.2w)}{i}
\fmfright{o}
\fmfshift{(0,-.2w)}{o}
\fmf{dbl_dashes_arrow, tension=1}{v1,o}
\fmf{dbl_dashes_arrow, tension=1}{i,v1}
\fmfv{d.sh=circle,l.d=0, d.f=empty,d.si=.3w,l=$1$}{v1}
\end{fmfgraph*}}=
\parbox{32mm}{
\begin{fmfgraph*}(30,20)
 \fmfset{arrow_len}{8}\fmfset{arrow_ang}{20}
\fmfleft{i}
\fmfright{o}
\fmf{dbl_dashes_arrow, tension=3}{i,v1}
\fmf{dbl_dashes_arrow, tension=3}{v2,o}
\fmf{dbl_wiggly,left,tension=1}{v1,v2}
\fmf{dbl_dashes_arrow,tension=1}{v1,v2}
\end{fmfgraph*}}+
\parbox{32mm}{
\begin{fmfgraph*}(30,20)
 \fmfset{arrow_len}{8}\fmfset{arrow_ang}{20}
\fmfleft{i}
\fmfright{o}
\fmf{dbl_dashes_arrow,tension=3}{i,v1}
\fmf{dbl_dashes_arrow, tension=3}{v2,o}
\fmf{dbl_plain_arrow,left,tension=1}{v1,v2}
\fmf{dbl_dots,left,tension=1}{v2,v1}
\end{fmfgraph*}}.
\end{align*}
The various $G$ functions may be reduced to one-loop forms as follows \cite{Grozin:2005yg},
\bsp
&G(1, 1, 1, 1, -1) = -\frac{1}{2} G(1, 1)^2, \\
&G(0, 1, 1, 1, 1) = G(1, 1) G(1, 1+\e),\\ 
&G(1, 1, 1, 1, 0) = G(1, 1)^2, \\
&G(1, 1, 1, 1, 1) = 
-\frac{1}{\e} (G(1, 1) G(2, 1) - G(1, 1) G(2, 1+\e)),\\ 
&G(0, 1, 1, 0, 1) = G(1, 1) G(1, \e), \\
&G(-1, 1, 1, 1, 1) = \frac{1}{2}G(1, 1)(G(1, 1 + \e) - G(1, \e)),\\ 
&G(-1, 1, 1, 0, 1) = \frac{1}{2}G(1, 1)(G(1, \e) - G(1, \e - 1)), \\
&G(0, 1, 1, -1, 1) = \frac{1}{2}G(1, 1)(G(1, \e) - G(1, \e - 1)),
\end{split}
\ee
where
\be
G(n_1,n_2) = \frac{\G(n_1+n_2-d/2) \G(d/2-n_1) \G(d/2-n_2)}
{\G(n_1)\G(n_2)\G(d-n_1-n_2)}, \qquad d=4-2\e.
\ee

\end{fmffile}
\bibliography{n2loops}
\end{document}